# Technical Memo: The Impact of Nancy Grace Roman Telescope's default image processing on the detectability of moving Solar system objects

Author: Joseph Masiero, Caltech/IPAC, jmasiero@ipac.caltech.edu

Version: 1.2

Revision date: 2025-09-24

## Context:

The Nancy Grace Roman Telescope is scheduled to launch in 2026 to conduct a wide-field survey of the sky at near-infrared wavelengths. Although Roman is unable to track objects moving at non-sidereal rates, there is recent interest in the potential capability of the telescope to support planetary defense by tracking and characterizing asteroids and comets (Holler et al, 2025, arXiv:2508.14412). However, the standard pipeline image processing scheme that the mission is planning to implement for the majority of its survey data will preferentially reject flux from all moving objects during the process of cosmic ray rejection. Here we describe the impact of the default Wide Field Imager (WFI) processing on moving object detection, and possible mitigations that could be employed to recover moving object observations. We do not discuss the potential application of the Roman Coronagraph to moving objects, as it is beyond the scope of this document.

## Background:

Roman, like JWST, uses detectors that are capable of sample-up-the-ramp (SUTR) readouts. This allows for non-destructive reads of the electron count on each pixel during each integration, with the final measurement of each pixel flux being the slope of the readouts over time, which directly corresponds to the brightness of the source on the pixel. SUTR readout schema have the benefit of being able to programmatically identify pixel saturation and cosmic ray events, and remove those samples before slope calculation, improving the quality of the resulting measurement. When backgrounds are low, SUTR is an optimal data collection choice, which is why it was chosen for most JWST observations. To obtain the maximum benefit from SUTR, it is imperative to record and downlink every sample taken, however this will significantly increase the data volume.

Due to downlink volume limitations and the large number of pixels in each detector, Roman is unable to send all samples to Earth. Instead, the project has opted for a multi-accumulation method, where a subset of readouts is averaged onboard the spacecraft, and only those "resultants" are downlinked (see Figure 1). Casertano (2022) provides a detailed discussion of the options for implementation of a multi-accumulation schema that provide balanced sensitivity to objects at a wide range of brightnesses. The pattern of the multi-accumulation is designed to optimize the use of the chosen exposure time, and can be changed depending on the survey being conducted.

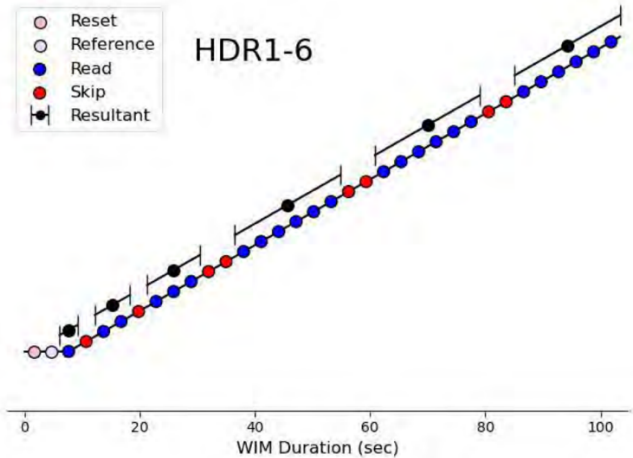

*Figure 1: An example of a possible multi-accumulation pattern that could be implemented by Roman. Here, each dot is a read out of the electrons on each pixel, with blue dots being averaged into the black 'resultants' that will be downlinked, and red points are skipped (not planned for typical Roman operations). The final derived slope calculated from the change in resultant value as a function of time acts as the measurements of the brightness of the source. Reproduced from Casertano (2022).*

By moving to resultants, Roman loses the ability to query individual read outs and identify flux-change outliers that would indicate cosmic ray strikes. Instead, the change between resultants would need to be compared, however the duration of each resultant and the point of time within each resultant that the cosmic ray hits the detector will change the appearance. Sharma & Casertano (2024) provide a detailed discussion of the method being used to identify jumps in the downlinked resultants. Figure 2 shows an example integration being hit by a cosmic ray, with perfect knowledge of each SUTR read, while Figure 3 shows how this would impact the difference between the downlinked resultants, which is the quantity that will be used during ground-processing to determine the slope and thus the brightness of the source. Notable here is that one cosmic ray will typically impact two resultant differences, requiring both to be dropped and marked as 'Jump' events in the image mask file that is produced during standard processing.

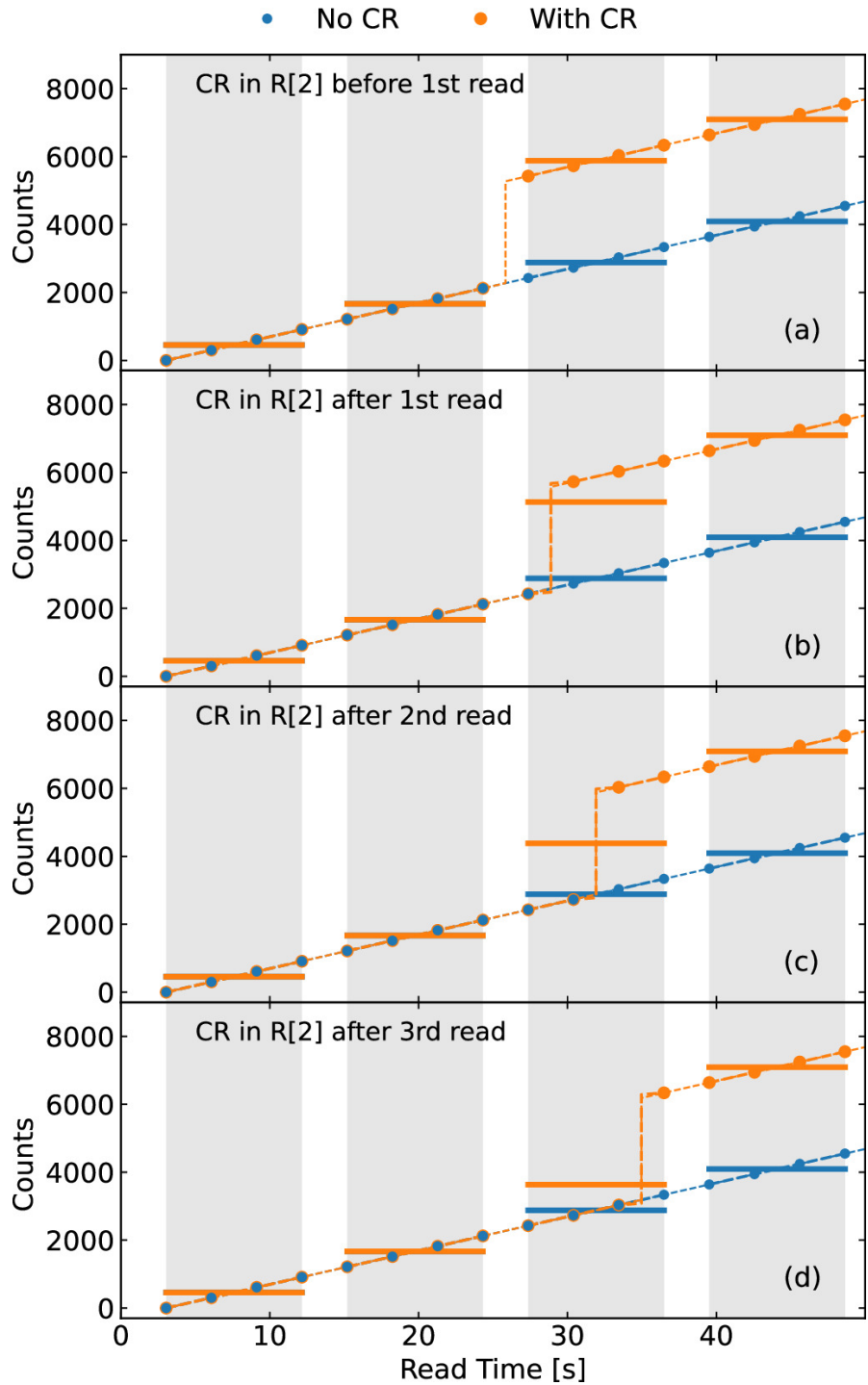

Figure 2: An example of the impact of transient sources (e.g. cosmic rays) on the resultants measured and downlinked by Roman depending on when the transient occurs. Reproduced from Sharma & Casertano (2024)

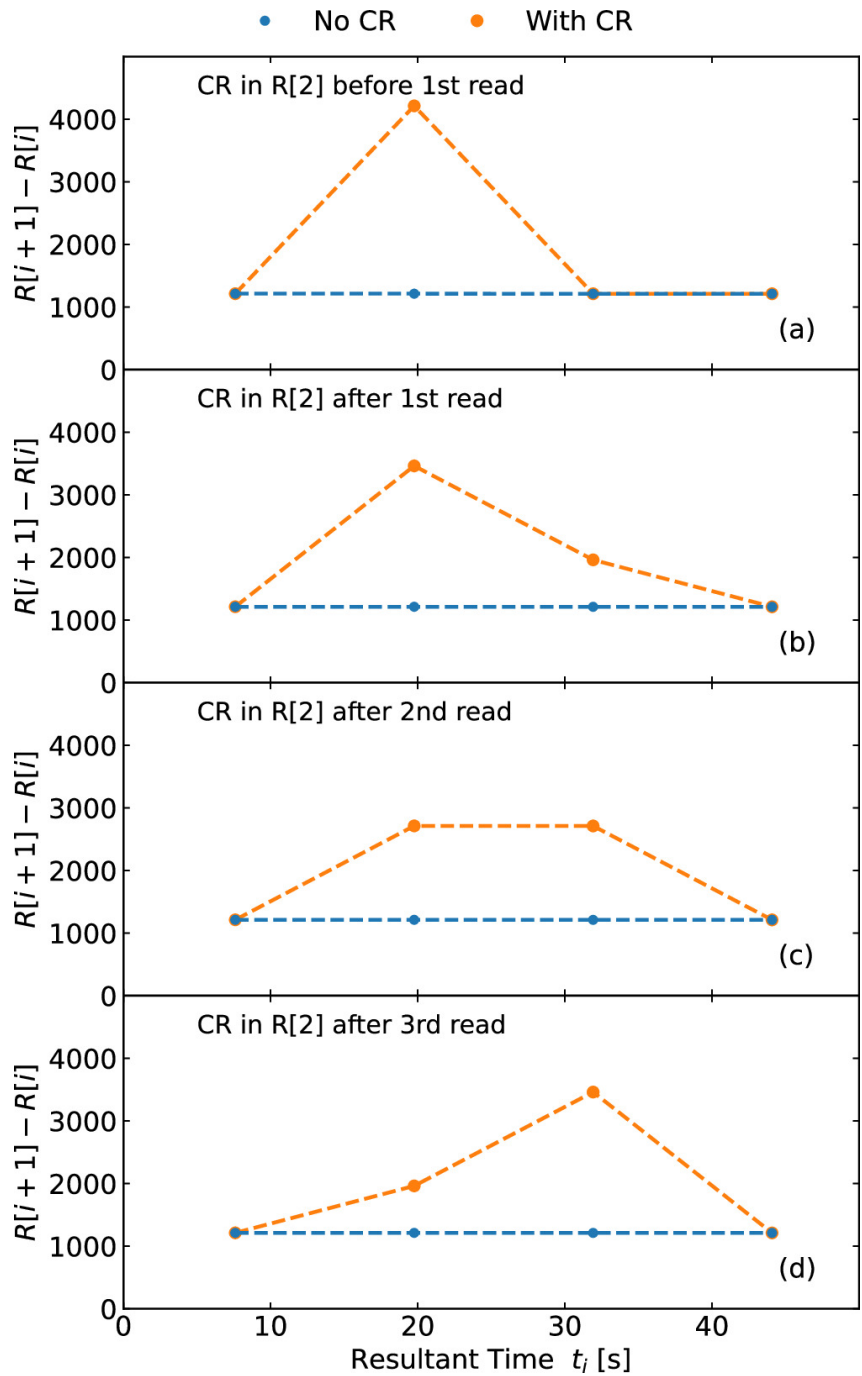

*Figure 3: The difference between successive resultants (representing the analyzed quantity on the ground) for the four different cosmic ray impact times within Resultant #2 shown in Fig 2. Reproduced from Sharma & Casertano (2024).*

### Effects on Moving Objects:

Two example multi-accumulation patterns from Casertano (2022) that will commonly be used during Roman survey imaging are:

HL300: [1, 1, 2, 2, 4] + [8] * 11

HL600: [1, 1, 2, 2, 4] + [8] * 6 + [11] + [16] * 8

In these cases, each read is a 3 second period, and the numbers indicate how many reads are incorporated in the downlinked resultants. HL300 is a total of 300 seconds of exposure time (made up of two 1-read resultants, two 2-read resultants, a 4-read resultant, and eleven 8-read resultants) while HL600 is a 600 second exposure. The initial short resultants are designed to give sensitivity to very bright objects that will saturate the full exposure, while the longer resultants provide better sensitivity to faint sources.

Roman has a very small pixel scale compared to most moving object surveys. The WFI detectors have a scale of 0.11" per pixel to make optimal use of the diffraction-limited optics provided by the space environment. Even objects as distant as the Kuiper Belt move ~2 arcsec/hour (0.01 deg per day), which corresponds to ~1 pixel of motion over a 300-sec exposure. NEOs have plane-of-sky motions that are two orders of magnitude greater, meaning that they will move ~1 pixel over the course of a 3-second read, and therefore many pixels over a full integration. For this reason, the flux from near-Earth objects will appear to the jump detection algorithm to be indistinguishable from a series of cosmic rays, and will be removed from the photometric calculation and flagged in the mask frame.

Therefore, moving objects can be identified by a cluster of jump flags within a frame, which will have a size and direction approximately matching of the object's on-sky motion. Due to the unbalanced resultant pattern, it is not a given that the flagging will be consistent, or complete across the entire streak of the object's motion. Likewise, the object's rate of motion will spread its flux over multiple pixels, diluting the sensitivity of the system.

An example of the effects of jump rejection on moving object flux is demonstrated by JWST data from Müller et al. (2023) and reproduced in Figure 4. In this case, asteroid (10920) was targeted by JWST for MIRI calibration images, which tracked at its predicted rate of motion during the observations. The data were processed through the standard image processing pipeline, which employs the same suite of routines that will be used for Roman image processing, including a very similar jump detection algorithm. JWST was tracking non-sidereally at the target's rate of motion, and so the object is recovered by the normal processing. However when looking at the raw SUTR reads that were downlinked by JWST, Müller et al. noted another source appeared to be in the frame. By turning off jump detection and shifting the SUTR reads to align with an arbitrary rate of motion, a very bright object was recovered in the frames that had been otherwise suppressed. As Roman nominally is downlinking resultants instead of the full SUTR readout sequence, this recovery method will not work for the majority of the survey data.

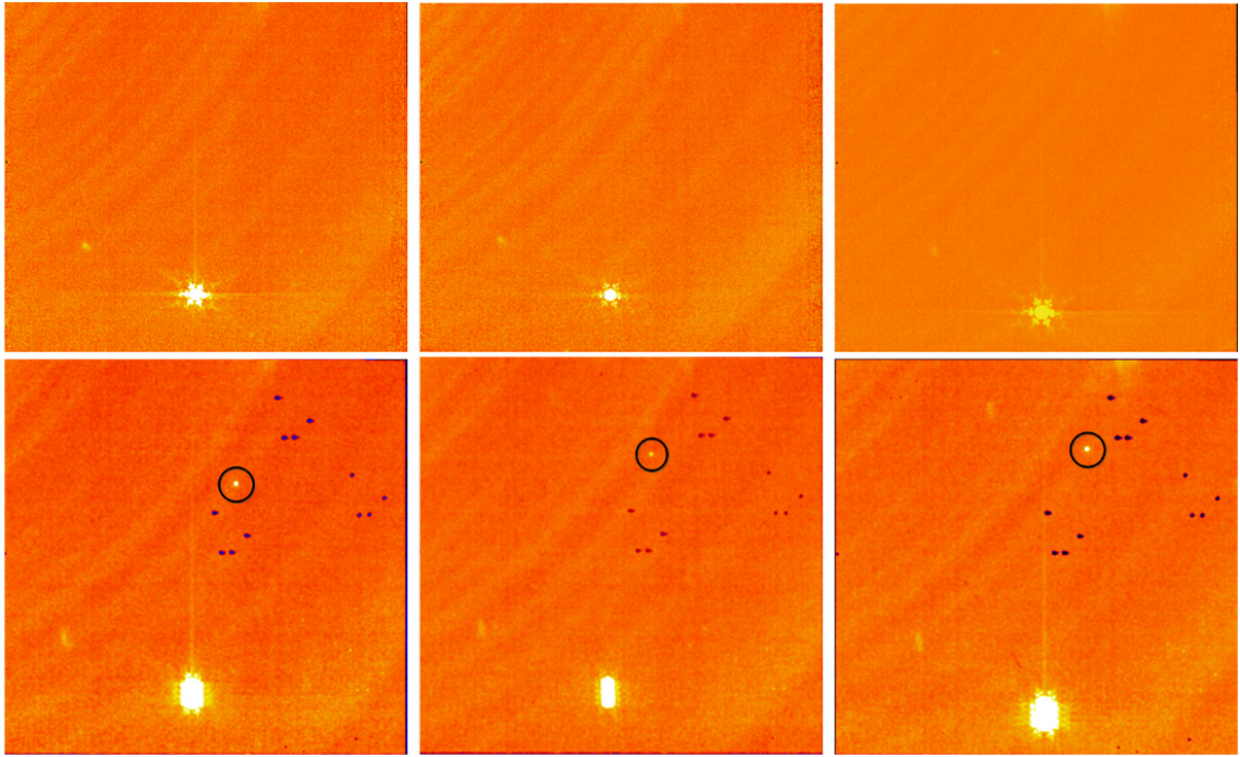

*Figure 4: The top row shows the JWST images of asteroid (10920) tracking non-sidereally on the object and using the default image pipeline processing, where the target asteroid is the bright source well-detected near the bottom of the frame, with the columns showing the filters F1000W, F1130W, and F1280W from left-to-right. The bottom row shows the same data, but reprocessed to recover objects moving with respect to the primary target, revealing a very bright (SNR>10) previously unknown asteroid which is circled in each image. The flux from this source was completely suppressed by the pixel jump detection/rejection algorithm in the top frames. Reproduced from Müller et al. (2023).*

## Potential Pathways Forward:

There are a few options that would make it possible for Roman to detect moving object sources, but each requires a change in operations to enable:

1. Downlink every read in the SUTR exposure pattern (this could be effected with a [1]*300 resultant pattern), turn off Jump detection in default processing, and search each image across the phase-space of possible moving objects; the negatives of this option are that it would dramatically increase downlink volume and processing time for these images, and would still suffer from dilution of faint sources due to trailing.
2. Limit observations to be a few seconds long so that moving objects appear static in each frame; like option 1 this will result in a data volume increase, loss of sensitivity to faint sources, and increased processing time.
3. Maintain the currently planned resultant pattern, but develop new algorithms to post-process images to remove the background and static source measured flux from jump-flagged resultants and attempt to convert the residual resultant counts to a flux of the underlying moving object; the negatives of this option are that both object detection and resultant-to-flux conversion would depend on a likely non-linear combination of the resultant pattern, the on-sky motion, and the target brightness, and this likely would only work for the brightest sources with faint objects not triggering the jump detection at all.
4. Enable non-sidereal tracking on Roman to allow for detection of both bright and faint moving objects with known rates of motion; the negatives of this option are that such a large change in operational capability at this late in the mission implementation would likely be infeasible without significant flight software rewrites and additional in-orbit checkout activities.

None of these options can be accomplished without some significant change in the current operations plan for Roman, which will have mission cost impacts and potentially schedule impacts as well.  Option 4 is the only one that ensures that the faintest sources can be recovered, and Options 1 and 2 would be the only ones that would potentially allow for new object discovery through the use of a shift-and-stack (aka synthetic tracking) post-processing of the short samples. Option 3 has the lowest impact on the default operations plan, but has not had a proof-of-concept demonstration to show that the JWST method of moving object recovery would work for the Roman resultants. Further, Option 3 is the most limited in the range of objects it would work for, and thus the hardest to estimate the total number of objects it could successfully recover photometry and/or astrometry for.

## Acknowledgements:

The RAPID project infrastructure team acknowledges NASA support under award 80NSSC24M0020 (program NNH22ZDA001N-ROMAN). JM thanks B. Holler, S. Carey, M. Egan, and R. Cosentino for discussion and comments.